\documentstyle[12pt]{article}
\author{Hans - J\"urgen Schmidt}
\title{Classical mechanics with lapse}
\date{}
\begin{document}
\maketitle

\bigskip

\centerline{
Universit\"at Potsdam, Institut f\"ur Mathematik,
Projektgruppe
Kosmologie}
\centerline{
      D-14415 POTSDAM, PF 601553, Am Neuen Palais 10, Germany}
\centerline{Tel. and Fax +49(0)331 962745;
   e-mail  hjschmi@rz.uni-potsdam.de}

\bigskip

\begin{abstract}
Mechanics  is  developed  over  a  differentiable
manifold  as space of possible positions.  Time is considered
to fill a one--dimensional Riemannian manifold,  so having the
metric as  lapse.  Then the system is quantized with covariant
instead of
partial derivatives in the Schr\"odinger operator.
\end{abstract}

\bigskip

\noindent PACS-numbers:  0320  (Classical  mechanics
of discrete  systems; general  mathematical  aspects);
0365 (Quantum  mechanics);  0240
(Differential geometry);  9880 (Cosmology).

\newpage

\noindent {\large{ \bf I. INTRODUCTION }}

\bigskip

In General Relativity, the differential quotient between
proper time and coordinate time is called lapse function. In
the present article, this notion is used for an arbitrary
classical mechanical system. Space is considered as
$n$--dimensional Riemannian space $V_n$ and time is
considered
as 1--dimensional Riemannian space $V_1$. Then the square of
the lapse function turns out to be the metric of this
$V_1$. Possible applications and comparison with other
approaches found in the literature will be shown in
section VI below.

Let  us consider a mechanical system.  The space of all
possible
positions shall be the $ n $-dimensional differentiable
manifold $ M_ n $.  It is endowed with local coordinates $ q
\sp i, i = 1, \dots n $. Most of all mechanical systems have
the property that
$ M_n$ is a subset of $ {\rm R} \sp m {\rm x } ( S \sp  1  )
\sp{n-m}  $,  so that the first $ m $ coordinates  are
Cartesian
ones and the remaining are periodic ones (i.e., angles). Here,
{\rm  R}  denotes  the  space of reals,  {\rm Z }  the  space
of integers,  and  the  one-dimensional torus  $ S \sp 1  $
can  be defined as factor space $ S \sp 1 = {\rm R} / {\rm Z }
$ . But in
general,  $ M  _n $ cannot be covered by one  single
coordinate
system.  The time is denoted by $ t $,  and $ \frac{d}{dt} $
will be  denoted by a dot.  So,  $ \dot q \sp i $ is the
velocity of a
moving  particle $ q \sp i (t) $. Therefore, the velocity at
time $ t $  is  an element of the tangent space  $ T _x M  _n
$ of  $  M  _n $ at $ x = q\sp i (t) $. The tangent bundle  $
TM _n $ is the union of all tangent spaces.

Contrarily  to  the usual procedure we now  introduce  the
lapse function  $ N(t) $ which shall be an arbitrary positive
function.
(Here   and  below  all  functions  shall  have   the
necessary
differentiability  properties.)  The  proper  time $  \tau  $
is defined by
\begin{equation}
\tau \  =  \ \int N(t) dt
\end{equation}
It  is uniquely determined up to an integration  constant,
i.e.,
without specifying the point where $ \tau = 0 $.
The  space  of  all possible times is a connected  oriented
one--dimensional Riemannian space $ V _1 $ with coordinate $ x
\sp 1 =
t  $  and metric $ g _{11} = N \sp 2 (t) $.  The  orientation
is chosen such that increasing time leads into the future.
So, Eq. (1)
represents the proper time $ \tau $ as proper length within
this $ V _1 $.

{ \it  Remark:  The definition is chosen such that  proper
time
does}  not  {\it  depend  on the velocity,  so we  do  not
cover
relativistic  effects. }

Each positive function $ N(t) $ defines a  gauge,  and
results  should not depend on it.  In this manner,  we define
the following gauge--invariant quantity, the proper velocity
$ v \sp i $
\begin{equation}
v \sp i \ =  \ \frac{1}{N} \ \dot q \sp i
\end{equation}
We have to prove that  $ v \sp i $ does not depend on the
special
choice of $N$; this follows from Eqs. (1, 2) via the equation
$$
v \sp i = \frac{dq \sp i}{d \tau }
$$
The action $ I $ is the integral of a Lagrangian $ L $
\begin{equation}
I \ =  \ \int L \ dt
\end{equation}
and  is  supposed to be a coordinate--,  gauge--,  and
 T--invariant
quantity. T--invariance  means
that  $I$ does not change if the orientation of  $V_1$  is
reversed.
 The  range  of integration in Eq. (3)  is  a  connected
subset of  $ V _ 1$, i.e., any fixed time--interval; but we do
not specify  now which kind of interval is used.

We  restrict ourselves to first--order Lagrangians,  i.e., $L$
is a function
\begin{equation}
L : TM _ n \ {\rm x } \ V _ 1 \longrightarrow {\rm R }
\end{equation}
The  next  three steps are done by plausible  arguments,  not
by proofs.

First, the explicit $t$--dependence
 ( $ t \in V _ 1 $ ) of $L $, Eq. (4),  is
 compatible with gauge--invariance of $ I $
 only  for the  case  that the $t$--dependence
  of $ L $ is via
$ N(t)  $  only, i.e.,
\begin{equation}
L \ = \ L( q \sp i ,  \dot q \sp i , N )
\end{equation}

Second, the coordinate-- and gauge--invariance of $ I $
requires the following form of $ L $
\begin{equation}
L \ = \ G ( q \sp i , v \sp i ) \cdot N
\end{equation}
where  $ G $ is a certain scalar;  this becomes  plausible
from Eqs. (1, 2, 3).

Third,  we  assume that $ G $ can be developed into powers of
$ v \sp i $
\begin{equation}
G \ = \ \sum _{k=0} \sp{ \infty } \alpha_{ i _1, \dots i _k }
\sp{(k)}
(q \sp i ) \ v \sp{i_1} \cdots v \sp{i_k}
\end{equation}
with  certain tensors  $ \alpha_{\dots} \sp{(k)} $.  Here,
and below,  the  Einstein sum convention is to be  applied.
Then  it
follows  from T--invariance,  that only even values $ k $ give
a non--vanishing contribution to Eq. (7).

The simplest non--trivial example for Eq. (7) is the case that
only
$  k = 0 $ and $ k = 2 $ give  contributions.  To meet the
usual notation we define
\begin{equation}
V \ = \ - \ \alpha \sp {(0) } ( q  \sp {i } ), \qquad
h_{ij} \ =  \ 2 \ \alpha_{ij} \sp {(2) } ( q \sp i )
\end{equation}
Inserting Eqs. (7, 8) into Eq. (6) we get
\begin{equation}
L \ = \ ( \frac{1}{2} h_{ij} v\sp i v \sp j - V ) \cdot N
\end{equation}
Without  loss  of  generality,  $  h_{ij} $ is assumed  to  be
a symmetric tensor in $ M _ n $. Here, the
coordinate--, gauge--, and
T--invariance of $ I $,  Eqs. (3, 9) is immediately seen; so
we also could have taken Eq. (9) as a definition of $ L $.

To give the Lagrangian Eq. (9) the structure defined by Eq.
(5)  we
insert Eq. (2) into Eq. (9) and get
\begin{equation}
L \  = \  \frac{1}{2} \ g_{ij} \ \dot q \sp i \ \dot q \sp j
\ - \ V  \cdot N
\end{equation}
where we used the definition
\begin{equation}
g_{ij} = \frac{1}{N} \cdot h_{ij}
\end{equation}

Next, we introduce the momentum $ p_i $ by
\begin{equation}
p _ i = \frac{ \partial L }{ \partial \dot q \sp i}
\end{equation}
{}From Eq. (10) we get
\begin{equation}
p _ i = g_{ij} \  \dot q \sp j
\end{equation}
It holds:  the momentum is gauge--invariant. This is proven by
the fact that from Eqs. (2, 11, 13) one gets
\begin{equation}
p _ i = h_{ij} \ v \sp j
\end{equation}
{}From Eqs. (12, 13) we get
\begin{equation}
\frac{ \partial \sp 2 L }{ \partial \dot q \sp  i \partial
\dot q
\sp j } \ = \  g _{ij}
\end{equation}
where  $  g _ {ij} $ depends on $ q \sp i $ and $ N $  only.
The analogous gauge--invariant equation to Eq. (15) reads
\begin{equation}
\frac{ \partial p _ i }{ \partial v \sp j } \ = \ h _{ij}
\end{equation}
and $ h _ {ij} $ depends on $ q \sp i $ only.

{ \it Remark:  One could use Eqs. (12, 15) also for the
general case $ L $,  Eqs. (6, 7); but then $ g_{ij} $ would in
general depend on the
velocities,  too.  If  $ g_{ij} $ is interpreted as metric,
then this would be the step from Riemannian to Finslerian
geometry.  A
typical example of Finslerian geometry appears,  if
the term with $ k = 4 $ in Eq. (7) is allowed to appear. }

Let us introduce the Hamiltonian
\begin{equation}
H \ = \ p _ i \ \dot q \sp i \  -  \ L
\end{equation}
The  canonical  equations make sense only for the case  that
the velocities  can  be  expressed as functions of  the
coordinates,
momenta, and time. Looking at Eq. (13) one can see that this
takes place  if and only if $ g _{ij} $ is a  regular  matrix.
So,  we
assume  this  to  be  the case in the following  and  denote
the inverse matrix to $ g _{ij} $ by $ g \sp {ij } $. From Eq.
(11) it
follows that also $ h _{ij} $ is invertible.
 The inverse matrix to $ h _{ij} $ is denoted
 by $ h  \sp {ij } $. It holds
\begin{equation}
g \sp {ij } = N \cdot h  \sp {ij }
\end{equation}
{}From Eq. (13) we get
\begin{equation}
\dot q \sp i = g \sp {ij }  \ p_j
\end{equation}
We insert Eqs. (10, 18, 19) into Eq. (17) and get
\begin{equation}
H \ = \ \frac{1}{2} \ g \sp {ij } \ p_i \ p_j  \ + \ V \cdot N
\end{equation}
which can also be written as
$H \ = \ (\frac{1}{2} \ h \sp {ij } \ p_i \ p_j \ +  \ V )
\cdot N $.  The canonical equations are
\begin{equation}
 \dot q \sp i = \frac{ \partial H}{ \partial p _ i}
\end{equation}
and
\begin{equation}
 \dot p _ i = - \frac{ \partial H}{ \partial q \sp i}
\end{equation}
Eq. (21) is equivalent to Eq. (19),  whereas Eq. (22)
represents the
equation  of motion;  in the next section we discuss it  in
more details.

\bigskip

\noindent {\large{ \bf II. THE EQUATION OF MOTION }}

\bigskip

The acceleration is $ a \sp i = \ddot q \sp i $.  In general,
the equation  of  motion  expresses the acceleration as
function  of coordinates, velocity, and time. To get this
structure, we insert
Eqs. (13, 20) into Eq. (22). After some calculus we get
\begin{equation}
a \sp i \ = \ \frac{ \dot N}{ N} \ \dot q^i \ - \ V  \sp {,i }
\cdot N \
- \ \dot q \sp  j \ \dot q \sp k \ \Gamma \sp i _ {jk}
\end{equation}
where $ V \sp{,i} = g \sp {ij } \ V _{,j} $ and  $ \Gamma $
denotes
the Christoffel affinity (which is the same both for $g _{ij}$
and $ h _{ij} $).  As usual,  $<< ,i>> $ is an abbreviation
for  the
partial derivative with respect to the coordinate $<< q \sp
i>> $.

We  can give three results immediately:  First,  for $N$ and
$V$ being  constant,  the  equation  of motion is just  the
geodesic equation  in the $ M _ n $ with Riemannian metric
$ g _ {ij} $.   Second,  for $N$ and $ g _{ij} $ being
 constant, the equation of
 motion reads $ 0 = a \sp i + V \sp{,i} $ and
equals the classical equation  of  motion in the  potential $
V$.  Third,  using  gauge--invariant quantities, we can write
the equation of motion as
\begin{equation}
0  =  \frac{ d v \sp i}{ d \tau}  + \Gamma \sp i _{jk} v \sp j
v \sp k + h \sp{ij} V _{,j}
\end{equation}
The  first  two  terms  of the  r.h.s.  represent  the
covariant
derivative of the proper velocity with respect to proper time.

In the next step we consider,  independently of the
Hamiltonian,
under  which  condition  the action $I$  Eq. (3)  has a
stationary value.  One  should  expect  that  the same
equation  of  motion
appears, but this is not fully trivial to show.

The corresponding Euler--Lagrange equation to the action $I$
reads
\begin{equation}
0 = \frac{\partial L}{ \partial q \sp i}
- \frac{d}{dt} ( \frac{\partial L}{ \partial \dot q \sp i} )
\end{equation}
With Eq. (12) we get
\begin{equation}
\dot p _i =  \frac{\partial L}{ \partial q \sp i}
\end{equation}
Comparing with Eq. (22) we have to show that
\begin{equation}
 \frac{\partial H}{ \partial q \sp i} = -
 \frac{\partial L}{ \partial q \sp i}
\end{equation}
Looking  at Eq. (17) one could get the impression that Eq.
(27) can be fulfilled for a constant product $ p _ i \dot q
\sp i $  only,
but  this  impression is wrong,  because in the l.h.s.,  $H$
is  a
function $H(q \sp i , p _ i , N ) $ but in the r.h.s., $L$ is
a function $ L  (q \sp i ,  \dot q \sp i ,  N ) $. And so,
with $H$
Eq. (20) and $L$ Eq. (10), the validity of Eq. (27) can be
proven.

\bigskip

\noindent {\large{ \bf III. THE LOWER--DIMENSIONAL CASES}}

\bigskip

Let  us  consider the simplifications for  the
lower--dimensional
cases.  For  $n = 1$,  one knows that the Riemannian space
$V_1$ is flat, and so the Lagrangian Eq. (9) reduces to
$ L = [ \frac{m}{2} v \sp 2 - V(x) ] \cdot N(t) $
with $ q \sp 1 = x, \  v \sp 1 = v $ and $ h _{11} = m =
const. \ne 0$.  With  $ N = 1 $ this is the usual  point
particle  in  a potential $V$.

For  $n = 2 $,  the Riemannian space $ V_2 = (M _2 , h_{ij}) $
need  not to be flat,  but it is always conformally flat.  So
one can always find local coordinates such that the Lagrangian
Eq. (9)  can be written as
\begin{equation}
 L = [ \frac{m}{2} v \sp 2 +  \frac{M}{2} w \sp 2 - W(x,y) ]
\cdot S(x,y) \cdot N(t)
\end{equation}
with $ q \sp 2 = y,  v \sp 2 = w $ and $ h _{22} = M = const.
\ne 0 $ and $ W \cdot S = V $ as additional relations.
$S \ne 0 $ is the suitably chosen conformal factor.

For $ n \ge 3 $,  however,  a $ V _n $ need not to be
conformally flat,  and so,  in general,  the usual kinetic
term with constant
masses can be reached neither by a coordinate nor by a
conformal transformation.

\bigskip

\noindent {\large{ \bf IV. QUANTIZATION }}

\bigskip

The  usual quantization procedure is to substitute $ p _ k $
by  $  i  \hbar  \frac{  \partial  }{ \partial q  \sp  k}  $
in  the
Hamiltonian to come from the function to the operator. If we
make this  in  our approach,  then gauge--invariance  is
automatically ensured,  because  both  $  q \sp k $ and $ p _
k  $  are  gauge--invariant   quantities.   (To   prevent
misunderstandings,   we
explicitly  say:  $ i $ is an index $ \in \{1,  \dots n \}  $
if written  in  index  position,   and  it  is  the  imaginary
unit otherwise.) But to ensure coordinate--invariance,  the
partial  derivative is not sufficient.  The most natural  way
to circumvent  this  difficulty is to use the  covariant
derivative with  the same $ \Gamma $ as before.  Then $ \nabla
_ k $ denotes
the covariant derivative with respect to  $ q \sp k $.

The world function is denoted by $ \psi $, it is a function
\begin{equation}
\psi : M _ n \longrightarrow { \rm C }
\end{equation}
where  { \rm C } denotes the set of complex numbers.

The  energy  of  the  system is $ E = H / N $.  It  is  a
gauge--invariant   scalar,   and   it  is   constant   along
classical trajectories:  $  \frac{dE}{dt} = 0 $ which follows
from Eqs. (20, 21, 22).

So  we  get the Schr\"odinger equation $ \hat H \psi = E
\cdot  N
\cdot \psi $ with $ \psi = \psi ( q \sp i ) $ and
\begin{equation}
\hat  H = - \frac{1}{2} \hbar \sp 2 g \sp {ij } \nabla _ i
\nabla
_ j + V \cdot N
\end{equation}
The zero energy Schr\"odinger equation simply reads
\begin{equation}
\hbar \sp 2 \Box \psi = 2 V \psi
\end{equation}
where  $  \Box $ denotes the D'Alembertian with  respect  to
the metric $ h _{ij} $,  i.e., $ \Box = h\sp{ij} \nabla _i
\nabla _ j $,  whereas  the  general Schr\"odinger equation
can be  obtained
from this one by a suitable redefinition of $ V $.

To  circumvent  the  explicit  calculation  of  the
Christoffel affi\-ni\-ties we apply the following formula
\begin{equation}
\Box \ = \ \frac{1}{ \sqrt{h}} \ \partial _ i \ \sqrt{h}
\ h \sp{ij} \ \partial _ j
\end{equation}
where $ h = \vert {\rm det} h _{ij} \vert \ne 0 $.

{\it Remark: One should observe that the form used here is
surely the  simplest   possible  way  to   get a
coordinate--invariant Schr\"odinger equation;  however, it is
not the only possible one
which  goes over to the classical Schr\"odinger  equation
(i.e., that one with partial derivatives) if $h _{ij} $
becomes
constant.   Indeed,  one  could  use  the  conformally
invariant
operator  $ \Box _c = \Box - \xi R $ instead of $ \Box  $,
where $R$ is the curvature scalar of the metric $ h _{ij} $
and $ \xi = \frac{n-2}{4(n-1)}$ . Only for $ n \le 2 $ one has
$\Box_ c = \Box $;  for $ n = 2 $ because of $ \xi = 0 $, and
for $ n = 1 $ because of $ R = 0 $.  But even for $ n \ge 3 $
one can
cover this variant by a suitable redefinition of $V$.   }

Let  us shortly say what happens for the lower--dimensional
cases.
For $ n = 1$,  one simply uses coordinates such that $ h _
{11} = 1 $ and one gets the usual equation.  For $ n = 2 $,
however, it
is a little more involved. We employ the fact that  $h_{ij} $
is conformally flat and so it can be written as
 $ h _ {ij} = \sqrt{h} \ \eta  _ {ij} $ where $  \eta  _ {ij}
$  is a  matrix in diagonal form where all diagonal elements
are $  \in \{ +1, -1 \} $. $ \eta \sp {ij} $ is the inverse to
$ \eta _ {ij} $; and, by construction, they coincide. Then we
insert Eq. (32) into Eq. (31) and get
\begin{equation}
\hbar \sp 2 \ \eta \sp {ij} \ \partial _ i \ \partial _ j
\ \psi \ = \ 2 \ \sqrt{h} \ V \ \psi
\end{equation}
The  l.h.s.  represents  the flat--space  D'Alembertian,  and
the factor $ \sqrt{h}$ in the r.h.s. can be absorbed by a
redefinition of $ V $.

For  $ n \ge 3 $,  however,  it requires special circumstances
to get  the  Schr\"o\-din\-ger  equa\-tion in the  form  of  a
flat--space D'Alem\-bertian.

\bigskip

\noindent {\large{ \bf V. SOLUTIONS OF THE
SCHR\"O\-DIN\-GER EQUA\-TION}}

\bigskip

{}From the full set of solutions of the
Schr\"odinger  equation  (31)  we are essentially  interested
in those  solutions which correspond to the classical
solutions  of the system (21,  22).  To this end we apply the
WKB-approximation and insert the ansatz
\begin{equation}
\psi = a \cdot \exp (iS/ \hbar )
\end{equation}
into Eq. (31) and get
\begin{equation}
\hbar \sp 2 \Box a + i \hbar ( 2 a _{,k} S \sp {,k} + a \Box S
) - a S _ {,k} S \sp {,k} = 2 a V
\end{equation}
where  $  S \sp {,k} = h \sp {jk }  S _ {,j}  $.  From Eq.
(34) we have  the situation that now two functions ( $a,  S$ )
represent one function ( $ \psi $ ).
So  we are free to put an additional relation as calibration.
It turns out that the following calibration is useful:  we set
for a moment  $  \hbar  = 0 $,  insert this into Eq. (35)  and
use  the resulting equation
\begin{equation}
S _{,k} S \sp {,k} + 2 V = 0
\end{equation}
as natural calibration. This is the usual classical limit.

Before  we  proceed  we  must  be  sure  that  Eq. (36)
possesses solutions.  If  the metric  $ h _{ij} $
 has indefinit  signature,
then this is trivial.  Let  $ h _{ij} $  be of definit
signature;
without loss of generality it shall be positively  definit,
for, otherwise,  simply $ V $ has to change its sign. In
regions where $ V \le 0 $, Eq. (36) has solutions, but in
regions with $ V > 0 $
it does not have any solutions. One should remember here, that
we have  redefined $ V $ such that the whole system has zero
energy.
So,  $  V  > 0 $ corresponds to a negative  kinetic  energy;
the latter  is impossible for a positively definit metric  $ h
_{ij}$.  We get as result:  the calibration Eq. (36) is
possible if and only if classical motion takes place there.

Now we insert Eq. (36) into Eq. (35) and get
\begin{equation}
0 \ = \ \hbar \ \Box a \ + \ 2 \ i \ a _{,k} \ S \sp {,k}
\ + \ i \ a \ \Box S
\end{equation}
To proceed, there exist different possibilities: first, one
again neglects the term with $ \hbar $,  second,  one requires
$ a $ to
be  a slowly varying amplitude such that $ \Box a $ is
negligible
in comparison with $ \Box S $,  or,  third, one thinks of $a$
and $S$  as  real  functions  and so Eq. (37) splits  into
real  and imaginary  parts.  It  is not so essential which of
these  three arguments  are  applied,  because all of them
give  rise  to  the equation
\begin{equation}
0 =  2  a _{,k} S \sp {,k} +  a \Box S
\end{equation}
Eq.  (38)  can  be solved as follows:  let  $ S(q \sp i) $  be
a solution of Eq. (36) with $ S _ {,k} \ne 0 $. There exists
no time in the system, but we can introduce a time $T$ by
requiring that
$ \frac{d}{dT} = S \sp {,k} \partial _ k $.  With $ b = \ln a
\sp 2 $, Eq. (38) now reads
\begin{equation}
\frac{db}{dT} = - \Box S
\end{equation}
which  can  be integrated along the trajectories of  $T$.  In
an afterwards--interpretation one can identify $T$ with $ \tau
$, $ S \sp  {,k} $ with $ v \sp k $ and $ S _{,k} $ with $ p
_k $;  this turns  out to be compatible with the classical ($
=$ non--quantum) equations.  But  this alone does not suffice:
from  Eq. (39)  one calculates  the function $ a( q \sp i ) $
and inserts it together with  $ S(q \sp i ) $ into Eq. (37).
Then  the  WKB--approximation
turns  out to yield results close to the exact solution only
for the case that indeed,  $ \vert \hbar \Box a \vert $ is
negligible
in  comparison  to $ \vert a \Box S \vert $.  So one can
 check  in which region the semiclassical approach
 makes sense.

\bigskip

\noindent {\large{ \bf VI. CONCLUSION }}

\bigskip

Classical mechanics, as is usually presented, e.g. in Refs. 1
and 2, uses essentially vector spaces as space of possible
positions. Then  one has the duality between coordinates and
momenta  (which we   do  loose  here)  and  can  build  a
symplectic   manifold.
Furthermore,  one has usually a constant mass tensor (which
means a  constant matrix $ h _{ij} $ in our notation).

 Both points are generalized  in  the  present  paper.  The
present  approach  is inspired by work on Hamiltonian quantum
cosmology,  e.g.  Ref. 3, where the space of possible
positions is the set of all  possible spatial geometries
(called superspace). The set of all possible
spatial geometries turns out to be neither a vector space nor
is the   matrix $h_{ij}$  a constant one.  Even, if one
restricts to  the  minisuperspace which corresponds to
homogeneous  spatial
geometries,  one does not get a vector space. Example: The set
of all homogeneous 3--spaces of Bianchi--type IX [i.e., there
exists a transitive  subgroup of  the  isometry group
isomorphic to  $ SO(3) $]  which is a manifold with boundary,
the interior is composed of points corresponding to spaces
whose isometry group is 3--dimensional, and the boundary
points are formed by spaces with 4--dimensional isometry group
(i.e., the axially symmetric Bianchi--type IX models), and the
edge (the boundary of the boundary) consists of one line which
itself corresponds to the isotropic 3--spheres with
6--dimensional isometry group. (Concerning details to this
point see e.g. Ref. 4).

If $M_n$ is such a manifold with boundary, then a trajectory
is simply mirrored at the boundary.

Here we carefully distinguish between co-- and contravariant
tensor indices, and the Einstein sum convention is used in its
strong version: summation over double indices takes place only
for the case that one of them is in upper ($=$ contravariant)
and the other in lower ($=$ covariant) position. It is a nice
additional check of the formulae that the necessity to write
the $\Sigma$--sign never appeared.

The essential result of the present paper is to show that up
to dimension two, the Schr\"odinger equation comes out with
the flat--space D'Alembertian whereas for higher dimensions,
it requires a special structure of the action to have
  this property. This has the following consequence
 for quantum
cosmology: All models with one-- or two--dimensional
minisuperspace can be written with the flat--space
D'Alembertian in the Schr\"odinger equation (which is called
Wheeler de Witt equation here), whereas for
higher--dimensional mini\-super\-space models, e.g. Ref. 5,
this property requires a special structure of the underlying
system.

Reformulated for the classical (i.e., non--quantized) system
one can state: A system with one or two degrees of freedom has
always a kinetic energy which can be written as sum of terms
of the type $ \ \pm \ \frac{m}{2} \ v^2 \ $
 with positive constant values $m$, whereas for three or
higher dimension this need not to be the case.

The kind of introducing the covariant derivative in Eq. (30)
instead of the partial one is the mathematical background of
the (today widely accepted) solution of the so--called
factor--ordering problem, which filled many papers on quantum
cosmology in the eighties, see Ref. 6 which is a bibliography
of papers on the topic.

We always wrote velocities with upper (contravariant) and
momenta with lower (covariant) index; this is more than
a purely notational arbitrariness, moreover, it is the only
adequate form from the differential geometric point of view.

A geometric description of non--relativistic quantum mechanics
has already carried out by Kucha\v r [7] in 1980. He uses a
degenerate metric (i.e., a metric with vanishing determinant),
so that he needs additional considerations to relate the
co-- and the contravariant components of it. He solves the
factor--ordering problem by writing the Laplacian covariant
with repect to this degenerate metric. Contrary to our
approach (see also Ref. [5] for more details), he uses Dirac's
constraint quantization.

Section 7.2 of Ref. [8] develops classical mechanics in
parametrized form. In this form, it becomes
time--reparametrization invariant just as General Relativity
is coordinate--invariant. Their approach takes velocities and
momenta on the same footing (both are covariantly
written vectors).

The book [9] by Zeh reviews many aspects of the
{\it direction} of time. In subsection 5.2.1 of that book,
also the reparametrization invariance of time is mentioned,
Zeh relates this property to Mach's principle
  (regarding time). Ref. [10] presents a geometrization
  of classical
mechanics by use of a symplectic structure.
 Refs. [11]
 discuss the recovering of time
and the deduction of the Wheeler de Witt equation in
 quantum cosmology.

An application of the present  approach (the present article
is a revised version of the unpublished Potsdam--Report No.
93/10 from January 1993) can be found in section V A
 of Ref. [12], where it is used to deduce the
Wheeler de Witt equation for the Starobinsky
 cosmological model. For further generalizations see e.g.
 Ref. [13].

\newpage

\noindent {\large {\bf Acknowledgment}}

\bigskip

I  thank Dr.  U.  Kasper,
Dr. M. Rainer, and the unknown referee
for valuable  comments.
Financial  support  from the
Wissenschaftler--Integrations--Programm
 contract 015373/E  and from Deutsche
 Forschungsgemeinschaft
contract Schm 911/5 are
gratefully acknowledged.

\bigskip

{\bf References }

\bigskip

\noindent 1.  L.  D.  Landau,  E. M. Lifshitz,
{\it Mekhanika} (in Russian, Izdatelstvo
Nauka,  Moscow  1957);  German translation: {\it Mechanik},
(Akademie--Verlag Berlin  1962);  English translation:
{\it Mechanics},  (Pergamon
Press London 1958).

\medskip

\noindent 2.  V.  I. Arnold, {\it Matematicheskie metody
klas\-siches\-koi mekha\-niki }
(in  Russian, Izd.  Nauka,   Moscow  1974);   English
 translation: {\it Mathematical  methods  of  classical
mechanics},   (Springer--Verlag
New York, Berlin 1978).

\medskip

\noindent 3.  M.  Ryan, {\it Hamiltonian Cosmology},
  (Springer--Verlag  New  York,
Berlin 1972).

\medskip

\noindent 4. H.-J. Schmidt, J. Math. Phys. {\bf 28}, 1928
(1987), {\bf 29}, 1264 (1988).

\medskip

\noindent 5. V. N. Lukash, H.-J. Schmidt, Astron. Nachr.
{\bf 309}, 25 (1988).

\medskip

\noindent 6. J. Halliwell, Int. J. Mod. Phys. A {\bf 5}, 2473
(1990).

\medskip

\noindent 7. K. Kucha\v r, Phys. Rev. D {\bf 22}, 1285 (1980).

\medskip

\noindent 8. R. Arnowitt, S. Deser, C. Misner, page 227 in:
 L. Witten (Ed.), {\it Gravitation: an introduction to current
research}, (Wiley New York 1962).

\medskip

\noindent 9. H. Zeh, {\it The physical basis of the direction
of time} (Springer--Verlag Berlin, New York 1992) 2nd ed.

\medskip

\noindent 10. F. Ghaboussi, J. Math. Phys. {\bf 34}, 4000
(1993).

\medskip

\noindent 11. C. Kiefer, Phys. Rev. D {\bf 47},
 5414 (1993); U. Kasper, Class. Quant. Grav. {\bf 10}, 869
(1993).

\medskip

\noindent 12. H.-J. Schmidt, Phys. Rev. D {\bf 49},
 6354 (1994).

\medskip

\noindent 13. M. Szydlowski, J. Szczesny, Phys. Rev. D {\bf
50}, 819 (1994).

\bigskip

\noindent Address of the author: \\
Dr. habil. Hans - J\"urgen Schmidt \\
 Universit\"at Potsdam \\
 Institut f\"ur Mathematik, Projektgruppe Kosmologie\\
 PF 601553
D - 14415 POTSDAM, Germany

\end{document}